\documentclass[a4paper,reqno,12pt]{amsart}
\usepackage[all]{xy}           %For commutative diagrams
\usepackage{amssymb}           %For double arrows
\usepackage{hyperref}
\usepackage{eucal}
\usepackage{epsfig}
\numberwithin{equation}{section}

\newtheorem{definition}{Definition}[section]
\newtheorem{lemma}[definition]{Lemma}
\newtheorem{theorem}[definition]{Theorem}
\newtheorem{proposition}[definition]{Proposition}

\newtheorem{remarkth}[definition]{Remark}

\newenvironment{remark}{\begin{remarkth}\upshape}{\hfill$\diamond$\end{remarkth}}

\renewcommand{\emph}[1]{{\bfseries\itshape{#1}}}

\newcommand{\R}{\mathbb{R}}      %Numeros reales
\makeatletter
\newcommand\prol{\@ifstar{\@proldf}{\@prolpf}}  %% if * dual else primal
\def\@prolpf{\@ifnextchar[{\@prolpf@wrt}{\@prolpf@}}
\def\@prolpf@wrt[#1]#2{\@ifnextchar[{\@prolpf@wrt@at{#1}{#2}}{\@prolpf@wrt@{#1}{#2}}}
\def\@prolpf@wrt@at#1#2[#3]{\prolsymbol^{#1}_{#3}#2}
\def\@prolpf@wrt@#1#2{\prolsymbol^{#1}#2}
\def\@prolpf@#1{\@ifnextchar[{\@prolpf@at{#1}}{\@prolpf@@{#1}}}
\def\@prolpf@at#1[#2]{\prolsymbol_{#2}#1}
\def\@prolpf@@#1{\prolsymbol#1}
\def\@proldf{\@ifnextchar[{\@proldf@wrt}{\@proldf@}}
\def\@proldf@wrt[#1]#2{\@ifnextchar[{\@proldf@wrt@at{#1}{#2}}{\@proldf@wrt@{#1}{#2}}}
\def\@proldf@wrt@at#1#2[#3]{\prolsymbol^{*#1}_{#3}#2}
\def\@proldf@wrt@#1#2{\prolsymbol^{*#1}#2}
\def\@proldf@#1{\@ifnextchar[{\@proldf@at{#1}}{\@proldf@@{#1}}}
\def\@proldf@at#1[#2]{\prolsymbol^*_{#2}#1}
\def\@proldf@@#1{\prolsymbol^*#1}
\def\prolsymbol{\mathcal{T}}
\makeatother

\begin{document}

\title[A universal Hamilton-Jacobi theory]{A universal Hamilton-Jacobi theory}
\author[M. de Le\'on]{Manuel de Le\'on}
\address{Manuel de Le\'on: Instituto de Ciencias Matem\'aticas (CSIC-UAM-UC3M-UCM),
c$\backslash$ Nicol\'as Cabrera, nº 13-15, Campus Cantoblanco,UAM
28049 Madrid, Spain} \email{mdeleon@icmat.es}

\author[D.\ Mart\'{\i}n de Diego]{David Mart\'{\i}n de Diego}
\address{David \ Mart\'{\i}n de Diego:
Instituto de Ciencias Matem\'aticas \linebreak (CSIC-UAM-UC3M-UCM),
c$\backslash$ Nicol\'as Cabrera, nº 13-15, Campus Cantoblanco,UAM
28049 Madrid, Spain}\email{david.martin@icmat.es}

\author[M. Vaquero]{Miguel Vaquero}
\address{Miguel Vaquero:
Instituto de Ciencias Matem\'aticas (CSIC-UAM-UC3M-UCM),
c$\backslash$ Nicol\'as Cabrera, nº 13-15, Campus Cantoblanco,UAM
28049 Madrid, Spain} \email{miguel.vaquero@icmat.es}


 \subjclass[2000]{}

\begin{abstract}
In this paper we develop a Hamilton-Jacobi theory in the setting of almost Poisson manifolds.
The theory extends the classical Hamilton-Jacobi theory and can be also applied to very general situations
including nonholonomic mechanical systems and time dependent systems with external forces.
\end{abstract}

\maketitle

\tableofcontents

\section{Introduction}

The standard formulation of the Hamilton-Jacobi problem is to find a
function $S(t, q^A)$ (called the {\bf principal function}) such that
\begin{equation}\label{hjj1}
\frac{\partial S}{\partial t} + h(q^A, \frac{\partial S}{\partial
q^A}) = 0,
\end{equation}
where $h=h(q^A, p_A)$ is the hamiltonian function of the system.
If we put $S(t, q^A) = W(q^A) - t E$, where $E$ is a constant, then
$W$ satisfies
\begin{equation}\label{hjj2}
h(q^A, \frac{\partial W}{\partial q^A}) = E;
\end{equation}
$W$ is called the {\bf characteristic function}.

Equations (\ref{hjj1}) and (\ref{hjj2}) are indistinctly referred as
the {\bf Hamilton-Jacobi equation} (see \cite{AM,Arnold,rund}).

The Hamilton-Jacobi equation helps to solve the Hamilton equations for $h$
\begin{equation}\label{ham}
\frac{dq^i}{dt} = \frac{\partial h}{\partial p_i} \; , \;
\frac{dp_i}{dt} = - \frac{\partial h}{\partial q^i}
\end{equation}
Indeed, if we find a solution $W$ of the Hamilton-Jacobi equation (\ref{hjj2}) then a solution $(q^i(t)$ of the first
set of equations (\ref{ham}) gives a solution of the Hamilton equations by taking
$p_i(t) = \frac{\partial W}{\partial q^i}$.

A geometric version of this result has been recently described by Cari\~nena {\it et al} \cite{cari},
based on the observation that if the hamiltonian vector field $X_h$ can be projected to the
configuration manifold by means of a 1-form $dW$ then the integral curves of the projected
vector field can be transformed into the integral curves of $X_h$ provided that $W$ is a solution of the Hamilton-Jacobi equation.

This observation has been succesfully applied to many other different contexts,
including nonholonomic mechanics (see \cite{cari,cari2,lmm1,lmm3}), singular lagrangian systems
\cite{hjsingular1,hjsingular2}, and even classical field theories \cite{lmm2,lmv,vilarino}.

The goal of the present paper is to present a general procedure for hamiltonian systems on
an almost-Poisson manifold, that is, a manifold equipped with a skew-symmetric $(2,0)$-tensor field
which does not necesarily satisfies the Jacobi identity. We also
assume that the almost-Poisson manifold has a fibered structure over another manifold.
The Hamilton-Jacobi problem now is to find a section of the fibered manifold
such that its image is a lagrangian submanifold and the differential of the given hamiltonian vanishes on
the tangent vectors to the section and belonging to the characteristic distribution.

The theory includes the case of classical hamiltonian systems on the cotangent bundle
of the configuration manifold as well as the case of nonholonomic mechanical systems. We also apply the theory to time-dependent hamiltonian systems and systems with external forces.

\section{Hamilton-Jacobi theory in almost-Poisson manifolds}

Let $\pi : E \longrightarrow M$ be a surjective submersion (in other words, a fibration) such that $E$ is equipped with an almost-Poisson
structure $\Lambda$, that is, $\Lambda$ is a skew-symmetric $(2,0)$-tensor field on $E$. Notice that $\Lambda$ does not necessarily satisfy
the Jacobi identy; in this case, we will have a Poisson tensor, and $E$ will be a Poisson manifold. For the moment, one only needs to
ask $(E, \Lambda)$ be an almost-Poisson manifold.

Therefore, $\Lambda$ defines a vector bundle morphism
$$
\sharp : T^*E \longrightarrow TE
$$
by
$$
\langle \sharp(\alpha), \beta \rangle = \Lambda (\alpha, \beta)
$$
for all 1-forms $\alpha$ and $\beta$ on $E$.

We denote by $\mathcal{C}$ the characteristic distribution defined by $\Lambda$, that is
\[
\mathcal{C}_p=\sharp(T_p^*E)
\]
for all $p \in E$.
The rank of the almost-Poisson structure at $p$ is the dimension of the space $\mathcal{C}_p$.
Notice that $\mathcal{C}$ is a generalized distribution and, moreover,  is not (in general) integrable
since $\Lambda$ is not Poisson in principle.

The following lemma will be useful
\begin{lemma}\label{lema} Let $(E,\Lambda)$ be an almost-Poisson manifold, then we have
\[
\mathcal{C}^{\circ}=ker(\sharp)
\]
\end{lemma}

{\bf Proof:}

Observe that
\begin{eqnarray*}
\left(\textrm{Im} \sharp_p\right)^{\circ}&=&\{\mu\in T^*_pE\; |\; \langle \mu, \sharp_p(\alpha)\rangle=0, \forall \alpha\in T_p^*E\}\\
&=&\{\mu\in T^*_pE\; |\; \langle \sharp_p(\mu), \alpha\rangle=0, \forall \alpha\in T_p^*E\}\\
&=&\ker\sharp_p
\end{eqnarray*}
and thus, the result holds.
\qed

We also have the following definition

\begin{definition}(\cite{liber,Vais})
A submanifold $N$ of $E$ is said to be a lagrangian submanifold if the following equality holds
\[
\sharp (TN^{\circ})=TN\cap \mathcal{C}
\]
\end{definition}

To have dynamics we need to introduce a hamiltonian function $h : E \longrightarrow \mathbb{R}$, and thus we obtain the corresponding hamiltonian vector field
$$
X_h = \sharp (dh).
$$

Assume that $\gamma$ is a section of $\pi : E \longrightarrow M$, i.e. $\pi \circ \gamma = id_M$. Define the vector field $X_h^\gamma$ on $M$ by
$$
X_h^\gamma = T \pi \circ X_h \circ \gamma
$$

The following diagram summarizes the above construction:

\[
\xymatrix{ E
\ar[dd]^{\pi} \ar[rrr]^{X_h}&   & &TE\ar[dd]^{T\pi}\\
  &  & &\\
 M\ar@/^2pc/[uu]^{\gamma}\ar[rrr]^{X_h^{\gamma}}&  & & TM }
\]

The following result relates the integral curves of $X_h$ and $X_h^\gamma$.

%\begin{remark}
%{\rm It should be noticed that we will
%indistinctly use $\gamma(M)$ or $\textrm{Im}(\gamma)$ to denote the image of a section $\gamma$.}
%\end{remark}

\begin{theorem}\label{master}
Assume that $\textrm{Im}(\gamma)$ is a lagrangian submanifold of $(E, \Lambda)$. Then the following assertions are equivalent:

\begin{enumerate}
\item $X_h$ and $X_h^\gamma$ are $\gamma$-related;

\item $dh\in (T\textrm{Im}(\gamma)\cap \mathcal{C})^{\circ}$.
\end{enumerate}
\end{theorem}

{\bf Proof:}

``$\Rightarrow$''

Assume that $X_h$ and $X_h^{\gamma}$ are $\gamma$-related. Then $X_h=T\gamma(X_h^{\gamma})$ and since $X_h\in\mathcal{C}$, we have $X_h\in T\textrm{Im}(\gamma)\cap \mathcal{C}$. But $\textrm{Im}(\gamma)$ is a lagrangian submanifold, so there exists $\beta\in(T\textrm{Im}(\gamma))^{\circ}$ such that 
\[
X_h=\sharp(\beta).
\]

Using that $X_h=\sharp(dh)$, we have $\sharp(dh-\beta)=0$, so $dh-\beta\in\textrm{Ker}(\Lambda)^{\sharp}=\mathcal{C}^{\circ}$.

Therefore
\[
\begin{array}{c}
dh\in\beta+\mathcal{C}^{\circ}\subset (T\textrm{Im}(\gamma))^{\circ}+\mathcal{C}^{\circ} \\ \noalign{\medskip}
=(T\textrm{Im}(\gamma)\cap\mathcal{C})^{\circ}.
\end{array}
\]
\bigskip

`$\Leftarrow$''

If $dh\in (T\textrm{Im}(\gamma)\cap \mathcal{C})^{\circ}=T\textrm{Im}(\gamma)^{\circ}+ \mathcal{C}^{\circ}$, then
$dh=\alpha_1+\alpha_2$ where $\alpha_1\in T\textrm{Im}(\gamma)^{\circ}$ and $\alpha_2\in \mathcal{C}^{\circ}$.

Then, along $\textrm{Im}(\gamma)$:
$$
X_h=X_{\alpha_1}+X_{\alpha_2}
$$
where $\Lambda(\alpha_i)=X_{\alpha_i}$, $i=1,2$. Using Lemma \ref{lema} we have $X_h=X_{\alpha_1}+X_{\alpha_2}=X_{\alpha_1}+0=X_{\alpha_1}$ where $\alpha_1\in T\textrm{Im}(\gamma)^{\circ}$.

Since $\textrm{Im}(\gamma)$ is a lagrangian submanifold, we have
\[
\sharp(T\textrm{Im}(\gamma)^{\circ})=T\textrm{Im}(\gamma)\cap \mathcal{C}
\]
and then
$$
X_h=X_{\alpha_1}\in T\textrm{Im}(\gamma)\cap \mathcal{C}
$$

Therefore we deduce that $X_h$ and $X^{\gamma}_{h}$ are $\gamma$-related since both are tangent to the section $\gamma(M)$.

\qed

Assume that $(E, \Lambda)$ is a transitive Poisson manifold, that is, $\mathcal{C} = TE$.
Then, we have
\begin{proposition}
A submanifold $N$ of $E$ is a lagrangian submanifold if and only if
\[
\sharp (TN^{\circ})=TN
\]
\end{proposition}

Therefore, the above theorem \ref{master} takes the following classical form.

\begin{theorem}\label{master2}
Assume that $\textrm{Im}(\gamma)$ is a lagrangian submanifold of $(E, \Lambda)$. Then the following assertions are equivalent:

\begin{enumerate}
\item $X_h$ and $X_h^\gamma$ are $\gamma$-related;

\item $d(h \circ \gamma) = 0$.
\end{enumerate}
\end{theorem}

\section{Computations in local coordinates}

Assume that $(x^i, y^a)$ are local coordinates adapted to the fibration $\pi : E \longrightarrow M$, that is,
$\pi(x^i, y^a) = (x^i)$, where $(x^i)$ are local coordinates in $M$.

Therefore, the tensor $\Lambda$ can be  locally expressed as follows
\begin{eqnarray*}
\Lambda &=& \frac{1}{2}\Lambda^{ij} \, \frac{\partial}{\partial x^i} \wedge \frac{\partial}{\partial x^j} +
\Lambda^{ib} \, \frac{\partial}{\partial x^i} \wedge \frac{\partial}{\partial y^b}
+ \frac{1}{2}\Lambda^{ab} \, \frac{\partial}{\partial y^a} \wedge \frac{\partial}{\partial y^b}
\end{eqnarray*}
where $\Lambda_{ij}=-\Lambda_{ji}$, $\Lambda_{ab}=-\Lambda_{ba}$ due to the antisymmetry of $\Lambda$.
Observe that
\begin{eqnarray*}
&& \Lambda^{ij} = \Lambda(dx^i, dx^j) \; , \; \Lambda^{ib} = \Lambda(dx^i, dy^b)\;,\\
&& -\Lambda^{ja} = \Lambda(dy^a, dx^j) \; , \; \Lambda^{ab} = \Lambda(dy^a, dy^b)\;.
\end{eqnarray*}

The above local expressions implies that
\begin{eqnarray}\label{local}
\sharp (dx^i) &=& \Lambda^{ij} \, \frac{\partial}{\partial x^j} + \Lambda^{ib} \, \frac{\partial}{\partial y^b} \\
\sharp (dy^a) &=& -\Lambda^{ja} \, \frac{\partial}{\partial x^j} + \Lambda^{ab} \, \frac{\partial}{\partial y^b}
\end{eqnarray}

Using \eqref{local} we deduce that a hamiltonian vector field $X_h$ for a hamiltonian function $h \in C^\infty(E)$ is locally expressed by
\begin{eqnarray}\label{hamup}
X_ h &=& (\frac{\partial h}{\partial x^i} \Lambda^{ij} - \frac{\partial h}{\partial y^a} \Lambda^{ja}) \, \frac{\partial}{\partial x^j}\\
&& + (\frac{\partial h}{\partial x^i} \Lambda^{ib} + \frac{\partial h}{\partial y^a} \Lambda^{ab}) \, \frac{\partial}{\partial y^b}
\end{eqnarray}

Now, let $\gamma : M \longrightarrow E$ be a section of $\pi : E \longrightarrow M$. If
$$
\gamma(x^i) = (x^i, \gamma^a(x^i))
$$
we obtain
\begin{eqnarray}\label{hamdown}
X_h ^\gamma = \left(\frac{\partial h}{\partial x^i} \Lambda^{ij} - \frac{\partial h}{\partial y^a} \Lambda^{ja}\right)\circ \gamma \, \frac{\partial}{\partial x^j}\; .
\end{eqnarray}

\begin{proposition}\label{master4}
$\textrm{Im}(\gamma)$ is a lagrangian submanifold of $(E, \Lambda)$ if and only if
\begin{equation}\label{laglocal}
\Lambda^{ab} - \Lambda^{jb} \frac{\partial \gamma^a}{\partial x^j} + \Lambda^{ja} \frac{\partial \gamma^b}{\partial x^j}
+ \Lambda^{ij} \frac{\partial \gamma^a}{\partial x^i} \frac{\partial \gamma^b}{\partial x^j} = 0
\end{equation}
\end{proposition}

{\bf Proof:}
First of all, let us observe that $T\textrm{Im}(\gamma)$ is locally generated by the
local vector fields
$$
\{ \frac{\partial}{\partial x^i} + \frac{\partial \gamma^a}{\partial x^i} \, \frac{\partial}{\partial y^a} \}
$$
since
$$
T\gamma(\frac{\partial}{\partial x^i}) =
\frac{\partial}{\partial x^i} + \frac{\partial \gamma^a}{\partial x^i} \, \frac{\partial}{\partial y^a}
$$
Therefore, if a 1-form
$$
\alpha =  \alpha_i dx^i + \alpha_a dy^a
$$
annihilates $T\textrm{Im}(\gamma)$ we deduce the following conditions on the coefficients:
\begin{equation}\label{coeficientes}
\alpha_ i = - \alpha_a \, \frac{\partial \gamma^a}{\partial x^i}
\end{equation}
Now, a simple computation shows that
$$
\sharp (\alpha) = (\alpha_i \Lambda^{ij} - \alpha_a \Lambda^{ja}) \, \frac{\partial}{\partial x^j}
+ (\alpha_i \Lambda^{ib} + \alpha_a \Lambda^{ab}) \, \frac{\partial}{\partial y^b}.
$$
Then, if $\sharp(\alpha) \in T\textrm{Im}(\gamma)$, with $\alpha \in T\gamma(M)^{\circ}$, and we use (\ref{coeficientes}) we deduce that
\begin{eqnarray*}
\sharp (\alpha) & = & (\alpha_i \Lambda^{ij} - \alpha_a \Lambda^{ja}) \, \frac{\partial}{\partial x^j}
+ (\alpha_i \Lambda^{ib} + \alpha_a \Lambda^{ab}) \, \frac{\partial}{\partial y^b} \\
&=& \alpha_a (- \frac{\partial \gamma^a}{\partial x^i} \Lambda^{ij} - \Lambda^{ja}) \,  \frac{\partial}{\partial x^i} +
\alpha_a (- \frac{\partial \gamma^a}{\partial x^i} \Lambda^{ib} + \Lambda^{ab}) \, \frac{\partial}{\partial y^b}\\
&=& \lambda^j  \, (\frac{\partial}{\partial x^j} + \frac{\partial \gamma^b}{\partial x^j} \frac{\partial}{\partial y^b})
\end{eqnarray*}
which implies
\begin{equation}\label{coef1}
\lambda^j = - \alpha_a \frac{\partial \gamma^a}{\partial x^i} \Lambda^{ij} - \alpha_a \Lambda^{ja}
\end{equation}
and
 \begin{equation}\label{coef2}
\lambda^j \frac{\partial \gamma^b}{\partial x^j} =
- \alpha_a \frac{\partial \gamma^a}{\partial x^i} \Lambda^{ib} + \alpha_a \Lambda^{ab}
\end{equation}
Substituting the values of $\lambda^j$ given by (\ref{coef1}) in equation (\ref{coef2}) we obtain
$$
\Lambda^{ab} - \Lambda^{jb} \frac{\partial \gamma^a}{\partial x^j} + \Lambda^{ja} \frac{\partial \gamma^b}{\partial x^j}
+ \Lambda^{ij} \frac{\partial \gamma^a}{\partial x^i} \frac{\partial \gamma^b}{\partial x^j} = 0.
$$

\hfill $\Box$

\section{Applications}

\subsection{Classical hamiltonian systems} (see \cite{AM,Arnold,deleonrodrigues})

A classical hamiltonian system is given by a hamiltonian function $h$ defined on the cotangent bundle
$T^*Q$ of the configuration manifold $Q$.

In this case, $E = T^*Q$ and $\Lambda$ is the canonical Poisson structure $\Lambda_Q$ on $T^*Q$ provided by the canonical symplectic
form $\omega_Q$ on $T^*Q$.
Recall that now we can take bundle coordinates $(q^i, p_i)$ where
$\pi_Q (q^i, p_i) = (q^i)$, and $\pi_Q : T^*Q \longrightarrow Q$ is the canonical projection.

Since in bundle coordinates
$$
\omega_Q = dq^i \wedge dp_i
$$
then
$$
\Lambda_Q  = \frac{\partial}{\partial q^i} \wedge \frac{\partial}{\partial p_i}
$$
Therefore,
$$
X_h = \frac{\partial h}{\partial p_i} \, \frac{\partial}{\partial q^i} - \frac{\partial h}{\partial q^i}\, \frac{\partial}{\partial p_i}
$$
and if a section $\gamma : Q \longrightarrow T^*Q$ (that is, a 1-form on $Q$) is locally expressed by
$$
\gamma(q^i) = (q^i, \gamma_i(q))
$$
we obtain
$$
X_h^\gamma = (\frac{\partial h}{\partial p_i}\circ \gamma) \, \frac{\partial}{\partial q^i}
$$

The notion of lagrangian submanifold defined in Section 2 in the almost-Poisson setting reduces to the well-known in the symplectic setting,
that is, it is isotropic and coisotropic with respect to the symplectic form $\omega_Q$.

If we compute the condition (\ref{laglocal}) in this case we obtain
$$
\frac{\partial \gamma_i}{\partial q^j} = \frac{\partial \gamma_j}{\partial q^i}
$$
which just means that $\gamma$ is a closed form, i.e., $d\gamma = 0$. So we recover the classical result (see \cite{AM,Arnold}).
\begin{proposition}
Given a 1-form $\gamma$ on $Q$, we have that $\textrm{Im}(\gamma)$ is a lagrangian submanifold
of $(T^*Q, \Lambda_Q)$ if and only if $\gamma$ is closed.
\end{proposition}

As a consequence, we deduce the classical result directly from Theorem \ref{master2}:
\begin{theorem}\label{master2-2}
Let $\gamma$ be a closed 1-form on $Q$. Then the following assertions are equivalent:

\begin{enumerate}
\item $X_h$ and $X_h^\gamma$ are $\gamma$-related;

\item $d(h \circ \gamma) = 0$.
\end{enumerate}
\end{theorem}

\subsection{Nonholonomic mechanical systems}

In this section we will recover the results obtained in two previous papers \cite{lmm1,lmm3} (see also \cite{cari2,sosa,bloch}).

A nonholonomic mechanical system is given by a lagrangian function $L : TQ \longrightarrow \mathbb{R}$
subject to contraints determined by a linear distribution $D$ on the configuration manifold $Q$.
We will denote by $\mathcal{D}$ the total space of the corresponding vector sub-bundle
$(\tau_Q)_{|\mathcal{D}} : \mathcal{D} \longrightarrow Q$ defined by $D$, where
$(\tau_Q)_{|\mathcal{D}}$ is the restriction of the canonical projection
$\tau_Q : TQ \longrightarrow Q$.

We will assume that the lagrangian $L$ is defined by a Riemannian metric $g$ on $Q$ and a potential energy $V \in C^{\infty}(Q)$, so that
$$
L(v_q) = \frac{1}{2} \, g(v_q, v_q) - V(q)
$$
or, in bundle coordinates $(q^i, \dot{q}^i)$
$$
L( q^i, \dot{q}^i) = \frac{1}{2} \, g_{ij} \dot{q}^i \dot{q}^j - V(q^i)
$$

If $\{\mu^a\}$, $1 \leq a \leq k$ is a local basis of the annihilator $D^o$ of $D$, then the constraints are
locally expressed as
$$
\mu^a_i (q) \, \dot{q}^i = 0,
$$
where $\mu^a= \mu^a_i(q) \, dq^i$.

The nonholonomic equations can be written as
\begin{eqnarray*}
&& \frac{d}{dt} \left( \frac{\partial L}{\partial \dot{q}^i} \right) - \frac{\partial L}{\partial q^i} = \lambda^i \mu^a_i(q)\\
&& \mu^a_i(q) \, \dot{q}^i = 0,
\end{eqnarray*}
for some Lagrange multipliers $\lambda^i$ to be determined.

Let $S$ (respectively, $\Delta$) be the canonical vertical endomorphism (respectively the Liouville vector field) on $TQ$. In local coordinates, we have
$$
S = dq^i \otimes \frac{\partial}{\partial \dot{q}^i} \; , \; \Delta = \dot{q}^i \, \frac{\partial}{\partial \dot{q}^i}
$$

Therefore, we can construct the Poincar\'e-Cartan 2-form $\omega_L $ \linebreak $=-S^*(dL)$ and the energy function function $E_L = \Delta(L) - L$, such that
the equation
\begin{equation}\label{symp}
i_{\xi_L} \, \omega_L = dE_L
\end{equation}
has  a unique solution, $\xi_L$, which is a SODE on $TQ$ (that is, $S(\xi_L)=\Delta)$. Furthermore,
its solutions coincide with the solutions of the Euler-Lagrange equations for $L$:
$$
\frac{d}{dt} \left( \frac{\partial L}{\partial \dot{q}^i} \right) - \frac{\partial L}{\partial q^i} = 0
$$

If we modify (\ref{symp}) as follows:
\begin{eqnarray}\label{symp-1}
i_{X} \, \omega_L - dE_L \in S^*((T \mathcal{D})^o)\\
X \in T\mathcal{D}
\end{eqnarray}
the unique solution $X_{nh}$ is again a SODE whose solutions are just the ones of the nonholonomic equations.

Let
$$
FL : TQ \longrightarrow T^*Q
$$
be the Legendre transformation given by
$$
FL (q^i, \dot{q}^i) = (q^i, p_i = \frac{\partial L}{\partial \dot{q}^i} = g_{ij} \dot{q}^j)
$$
$FL$ is a global diffeomorphism which permits to reinterpret the nonholonomic mechanical system in the hamiltonian side.
Indeed, we denote by $h = E_L \circ FL^{-1}$ the hamiltonian function and by $M = FL(\mathcal{D})$ the constraint submanifold
of $T^*Q$.

The nonholonomic equations are then given by
\begin{eqnarray*}
&& \frac{dq^i}{dt}  = \frac{\partial h}{\partial p_i}\\
&&  \frac{dp_i}{dt} = - \frac{\partial h}{\partial q^i} + \bar{\lambda}^i \mu^a_i,
\end{eqnarray*}
where $\bar{\lambda}^i$ are new Lagrange multipliers to be determined.

As above, the symplectic equation
$$
i_{X_h} \, \omega_Q = dh
$$
which gives the hamiltonian vector field $X_h$ should be modified as follows to take into account the nonholonomic constraints:
\begin{eqnarray}
i_{X} \, \omega_Q - dh \in F^o  \label{symp21}\\
X \in TM \label{symp22}
\end{eqnarray}
where $F$ is a distribution along $M$ whose annihilator $F^o$ is obtained from $S^*((T \mathcal{D})^o)$ through $FL$.
Equations (\ref{symp21}) and (\ref{symp22}) have a unique solution, the nonholonomic vector field $\overline{X}_{nh}$.

An alternative way to obtain $\overline{X}_{nh}$ is to consider the Whitney sum decomposition
$$
T(T^*Q)_{|M} = TM \oplus F^\perp
$$
where the complement is taken with respect to $\omega_Q$.
If
$$
P : T(T^*Q)_{|M} \longrightarrow TM
$$
is the canonical projection onto the first factor, one easily proves that
$$
\overline{X}_{nh} = P(X_h)
$$

Moreover, one can introduce an almost-Poisson tensor $\Lambda_{nh}$ on $M$ by
$$
\Lambda_{nh} (\alpha, \beta) = \Lambda_Q(P^* \alpha, P^*\beta)
$$
which is called the nonholonomic bracket (see \cite{nonlinear}).

Obviously, we have
$$
\overline{X}_{nh} = \sharp_{nh} (dh)
$$

\bigskip

An alternative way to define the nonholonomic bracket is as follows.
Consider the distribution
$$
TM\cap F
$$
along $M$. A direct computation shows that the subspace
$$
T_pM \cap F_p
$$
is symplectic within the symplectic vector space $(T_p(T^*Q), \omega_Q(p))$, for all $p \in M$ (see \cite{bates,nonlinear}).

Thus we have a second Whitney sum decomposition
$$
T(T^*Q)_{|M} = (TM \cap F) \oplus (TM \cap F)^\perp
$$
where the complement is taken with respect to $\omega_Q$.

If
$$
\tilde{P}: T(T^*Q)_{|M} \longrightarrow TM \cap F
$$
is the canonical projection onto the first factor, one easily proves that
$$
\overline{X}_{nh} = \tilde{P}(X_h)
$$

Moreover, it is possible to write $\Lambda_{nh}$ in terms of the projection $\tilde{P}$ as follows the the nonholonomic almost-Poisson tensor$\Lambda_{nh}$ on $M$ is now rewritten as
$$
\Lambda_{nh} (\alpha, \beta) = \Lambda_Q(\tilde{P}^* \alpha, \tilde{P}^*\beta)=\omega_Q(\tilde{P}(X_{\alpha}), \tilde{P}(X_{\beta}))
$$
(see \cite{nonlinear} for a proof).

Consider now the fibration
\[
\xymatrix{
(M, \Lambda_{nh}) \ar[d]^{{\pi_Q}_{|M}}\\
Q
}
\]
and the hamiltonian $h_{|M}$ (also denoted by $h$ for sake of simplicity).

We can easily prove that
\[
\mathcal{C}_p=T_pM\cap F_p
\]
Indeed,we have
\begin{eqnarray*}
\langle \sharp_{nh}(\alpha), \beta \rangle &=& -\omega_Q( \tilde{P}X_\alpha, X_\beta) =
 \omega_Q (X_{\beta}, \tilde{P}X_\alpha) \\
&=& (i_{X_\beta} \, \omega_Q) (\tilde{P}X_\alpha) =
 \langle \beta, \tilde{P}X_\alpha \rangle
\end{eqnarray*}
which implies
$$
 \sharp_{nh}(\alpha) =   \tilde{P}(X_\alpha)
$$

Furthermore, the symplectic structure $\Omega_p$ on $\mathcal{C}_p$ at any point $p \in M$ is given by
the restriction of the canonical symplectic structure $\omega_Q$ on $T^*Q$ to $\mathcal{C}_p$.

\begin{proposition}\label{global}
Let $\gamma:Q\rightarrow M$ be a section of ${\pi_Q}_{|M}:M\longrightarrow Q$, then $\textrm{Im}(\gamma)$ is a lagrangian submanifold of $(M, \Lambda_{nh})$
if and only if $d\gamma(X,Y)=0$ for all $X, Y\in D$.
\end{proposition}

{\bf Proof:}
We notice that $F=\{v\in T(T^*Q)\textrm{ such that } T\pi_Q(v)\in D\}$ and an
easy computation in local coordinates shows that dim$(F\cap TM)=2\,$dim($D$). Thus, we have
\[T\textrm{Im}(\gamma)\cap \mathcal{C}=T\gamma(D)\]

On the other hand, it is clear that our definition of lagrangian submanifold is equivalent to  $T\textrm{Im}(\gamma)\cap \mathcal{C}$
be lagrangian with respect to the simplectic structure $\Omega$ on the vector space $\mathcal{C}$.
Since $\Omega$ is the restriction of $\omega_Q$, given $X,\ Y \in D$ we have
\[
\Omega(T\gamma(X),T\gamma(Y))=\Omega_Q(T\gamma(X),T\gamma(Y))=d\gamma(X,Y)
\]

So, after a careful counting of dimensions, we deduce that
Im$(\gamma)$ is lagrangian with respect to $\Lambda_{nh}$ if and only if
$d\gamma(X,Y)=0$ for all $X,\ Y \in D$.

\qed

Using this proposition we can recover the Nonholonomic Hamilton-Jacobi Theorem as a consequence of Theorem \ref{master} (see \cite{lmm1,lmm3,bloch}).

\begin{theorem}\label{master5}[Nonholonomic Hamilton-Jacobi] Given a hamiltonian $h:M\rightarrow \mathbb{R}$, and $\gamma$ a $1$-from on $Q$ taking values in $M$, such that $d\gamma(X,Y)=0$ for all $X,\ Y \in D$, then the following conditions are equivalent
\begin{enumerate}
\item $\overline{X}_{nh}$ and $\overline{X}_{nh}^{\gamma}$ are $\gamma$-related.
\item $dh\in (T\gamma(D))^{\circ}$ (which is in turns equivalent to $d(h\circ\gamma)\in D^{\circ}$).
\end{enumerate}
\end{theorem}

\bigskip

We will get a suitable expression for the nonholonomic bracket $\Lambda_{nh}$ defined on the
constraint submanifold $M$ of $T^*Q$ (we follow the notations in \cite{nonlinear}). This local representation
can be also used to prove Proposition \ref{global}.

Let us recall that the constraints were defined through a distribution $D$ on $Q$.
Let $D'$ a complementary distribution of $D$ in $TQ$ and assume that
$\{X_\alpha\}$, $1 \leq \alpha \leq n-k$ is a local basis of $D$ and that
$\{Y_a\}$, $1 \leq a \leq k$ is a local basis of $D'$.  Notice that
$$
\mu^a (X_\alpha) = 0.
$$

Next we introduce new coordinates in $T^*Q$ as follows:
$$
\tilde{p}_{\alpha} = X^i_\alpha p_i \; , \tilde{p}_{n-k+a} = Y^i_a p_i
$$
where
$$
X_\alpha = X_\alpha^i \, \frac{\partial}{\partial q^i} \; , \;
Y_a = Y_a^i \, \frac{\partial}{\partial q^i}
$$

In these new coordinates we deduce that the constraints become
$$
\tilde{p}_ {n-k+a} = 0
$$
Therefore, we can take local coordinates $(q^i, \tilde{p}_\alpha)$ on $M$.

A direct computation shows now that the nonholonomic bracket $\Lambda_{nh}$ on $M$ is given by \cite{nonlinear}
\[
\begin{array}{ll}
\Lambda_{nh}(dq^i, dq^j) = 0 \; , \Lambda_{nh}(dq^i, d\tilde{p}_\alpha )= X^i_\alpha \\ \noalign{\medskip}
\Lambda_{nh}(d\tilde{p}_\alpha, d\tilde{p}_\beta) = X^i_\beta p_j \frac{\partial X^j_\beta}{\partial q^i} -
X^i_\alpha p_j \frac{\partial X^j_\beta}{\partial q^i}
\end{array}
\]

In the sequel, we will apply then general theory developed in Section 2 to the
almost-Poisson structure $(M, \Lambda_{nh})$.

Assume that $\gamma : Q \longrightarrow M$ is a section of $\pi : M \longrightarrow Q$. Then, we have
$$
\gamma (q^i) = (q^i, \tilde{\gamma}_\alpha (q^i))
$$
Since $\gamma$ can also be considered as a 1-form on $Q$ taking values on $M$ we have
$$
\gamma(q^i) = (q^i), \gamma_i(q^i))
$$
and since it takes values in $M$ we get
$$
\tilde{\gamma}_\alpha = X^i_\alpha \, \gamma_i
$$

A direct computation from equation (\ref{laglocal}) gives
\begin{eqnarray*}
 && \Lambda_{nh}^{\alpha \beta} + \Lambda_{nh}^{\beta j} \, \frac{\partial \tilde{\gamma}_\alpha}{\partial q^j}
-  \Lambda_{nh}^{\alpha j} \, \frac{\partial \tilde{\gamma}_\beta}{\partial q^j} \\
&& = X^i_\beta \gamma_j \frac{\partial X^j_\alpha}{\partial q^i} - X^i_\alpha \gamma_j \frac{\partial X^j_\beta}{\partial q^i}
- X^j_\beta \frac{\partial}{\partial q^j} \left(X^i_\alpha \gamma_i\right)
- X^j_\alpha \frac{\partial}{\partial q^j} \left(X^i_\beta \gamma_i\right)\\
&& = X^i_\alpha X^j_\beta \left(\frac{\partial \gamma_j}{\partial q^i} - \frac{\partial \gamma_i}{\partial q^j} \right) = 0.
\end{eqnarray*}
which can be equivalently written as
\begin{equation}\label{ideal}
d\gamma (X_\alpha, X_\beta) = 0
\end{equation}

Therefore, $\gamma(Q)$ is a lagrangian submanifold
of $(M, \Lambda_{nh})$ if and if $d\gamma \in \mathcal{I}(D^o)$, where
$\mathcal{I}(D^o)$ denotes the ideal of forms generated by $D^o$.
Indeed, notice that (\ref{ideal}) holds if and only if $d\gamma = \sum_a \, \xi_a \wedge \mu^a$, for some 1-forms $\xi_a$.

\subsection{Time dependent systems}
In this section we will follow \cite{deleonrodrigues}.
We can also develop a time-dependent version of the previous construction. If we have the fibration $E\rightarrow M$ such that $E$ is equipped with an almost-Poisson structure $\Lambda$, we can construct the following fibration in the obvious way
\begin{equation}\label{dia}
\xymatrix{
\mathbb{R}\times E\ar[d]^{\pi_{\mathbb{R}}}\\
\mathbb{R}\times M
}
\end{equation}
where now $\mathbb{R}\times E$ is equiped with the almost-Poisson structure given by the addition of the null bivector on $\mathbb{R}$ and $\Lambda$ on $E$.

We can consider the ``extended'' version of this diagram, that is, consider $T^*\mathbb{R}\times E$, equipped with the almost-Poisson structure  $\Lambda_{ext}$ given by the addition of the canonical Poisson structure on $T^*\mathbb{R}$ and $\Lambda$. Notice that if we consider coordinates global $(t,e)$ on $T^*\mathbb{R}\cong \mathbb{R}\times \mathbb{R}$, then
\[
\Lambda_{ext}=\frac{\partial}{\partial t}\wedge\frac{\partial}{\partial e}+\Lambda
\]
the canonical projection is
\begin{equation}\label{mu}
\begin{array}{rccl}
\mu: & T^*\mathbb{R}\times E &\longrightarrow & \mathbb{R}\times E \\ \noalign{\medskip}
&(t,e,p)& \rightarrow & \mu(t,e,p)=(t,p)
\end{array}
\end{equation}

According to the above notation, diagram \eqref{dia} becomes
\[
\xymatrix{
& T^*\mathbb{R}\times E \ar[dd]^{\tilde{\pi}} \ar[dl]_{\mu} \\
\mathbb{R}\times E\ar[dr]^{\pi_{\mathbb{R}}} \\
&\mathbb{R}\times M
}
\]
where $\tilde{\pi}=\pi_{\mathbb{R}}\circ\mu$.

Given a time dependent hamiltonian $h:\mathbb{R}\times E\rightarrow \mathbb{R}$, the dynamics are given by the evolution vector field $\frac{\partial}{\partial t}+X_h\in \mathfrak{X}(\mathbb{R}\times E)$. We can introduce the extended hamiltonian $h_{ext}:T^*{\mathbb{R}}\times E\rightarrow \mathbb{R}$ given by $h_{ext}=\mu^*h+e$ and the respective hamiltonian vector field $X_{h_{ext}}=\Lambda_{ext}^{\sharp}(dh_{ext})$. Notice that $\mu_*(X_{h_{ext}})=\frac{\partial}{\partial t}+X_h$.

We will denote by $\mathcal{C}_{ext}$ the characteristic distribution of $\Lambda_{ext}$. Notice that $\mathcal{C}_{ext}(t,e,p)=\langle \frac{\partial }{\partial t},  \frac{\partial }{\partial e}\rangle+ \mathcal{C}_p$, under the obvious identifications.

If $\gamma$ is a section of $\tilde{\pi}$, we can consider the section of $\pi_{\mathbb{R}}$ given by $\mu\circ\gamma$ and define the vector field $(\frac{\partial}{\partial t}+X_h)^{\gamma}$ on $\mathbb{R}\times M$ as follows:
\[
(\frac{\partial}{\partial t}+X_h)^{\gamma}=T\pi_{\mathbb{R}}\circ (\frac{\partial}{\partial t}+X_h)\circ (\mu\circ \gamma)
\]

Now, we can state the time-dependent version of the Hamilton-Jacobi theorem.

\begin{theorem}\label{teo}If Im($\gamma$) is a lagrangian manifold in $(T^*\mathbb{R}\times E,\Lambda_{ext})$, then the following assertions are equivalent.
\begin{enumerate}
\item $(\frac{\partial}{\partial t}+X_h)$ and  $(\frac{\partial}{\partial t}+X_h)^{\gamma}$ are $\mu\circ\gamma$-related
\item $dh_{ext}\in (T\textrm{Im}(\gamma)\cap \mathcal{C}_{ext})^{\circ}+\langle dt\rangle$
\end{enumerate}
\end{theorem}
{\bf Proof:}

``$\Rightarrow$''

Assume that $(\frac{\partial}{\partial t}+X_h)$ and  $(\frac{\partial}{\partial t}+X_h)^{\gamma}$ are $\mu\circ\gamma$-related. This means that given $m\in M$
\[
T\mu\circ T\gamma((\frac{\partial}{\partial t}+X_h)^{\gamma}(m))=(\frac{\partial}{\partial t}+X_h)(\mu\circ\gamma(m))
\]
or equivalently, there exists  $B\in\mathbb{R}$ such that
\[
T\gamma((\frac{\partial}{\partial t}+X_h)^{\gamma}(m))=(X_{h_{ext}}+B\frac{\partial}{\partial e})(\gamma(m))
\]
since any tangent vector in $T_{\gamma(m)}(T^*\mathbb{R}\times E)$ wich projects by $\mu$ onto $\frac{\partial}{\partial t}+X_h$ is of the form
\[
X_{h_{ext}}+B\frac{\partial}{\partial e},\quad B\in\mathbb{R}.
\]

Using the same argument that we used in Theorem \ref{master} we can conclude that
\[
dh_{ext}(\gamma(m))+Bdt\in (T_{\gamma(m)}\textrm{Im}(\gamma)\cap \mathcal{C}_{ext}(\gamma(m)))^{\circ}
\]
and so
\[
dh_{ext}\in (T\textrm{Im}(\gamma)\cap \mathcal{C}_{ext})^{\circ}+\langle dt\rangle
\]

``$\Leftarrow$'' 

Assume that $dh_{ext}\in (T\textrm{Im}(\gamma)\cap \mathcal{C}_{ext})^{\circ}+\langle dt\rangle$; this means that given any point $u\in\textrm{Im}(\gamma)$, there exists a real number $B$ such that 
\[
dh_{ext}(u)+Bdt(u)\in (T_x\textrm{Im}(\gamma)\cap (\mathcal{C}_{ext})_u)^{\circ}.
\]
Now we can deduce
\[
\Lambda_{ext}^{\sharp}(dh_{ext}(u)+Bdt(u))\in T_u\textrm{Im}(\gamma)
\]
where $\Lambda_{ext}^{\sharp}(dh_{ext}(u)+Bdt(u))=X_{h_{ext}}(u)+B\frac{\partial}{\partial e}(u)$

Obviously, the last statement implies that $T\mu_*(X_{h_{ext}}(u)+B\frac{\partial}{\partial e}(u))\in T_{\mu(x)}\textrm{Im}(\mu\circ\gamma)$, but
\[
\begin{array}{ll}
T\mu_*(X_{h_{ext}}(u)+B\frac{\partial}{\partial e}(u))&=T\mu_*(X_{h_{ext}}(u))+T\mu_*(B\frac{\partial}{\partial e}(u))\\\noalign{\medskip}&=T\mu_*(X_{h_{ext}}(u)) =(\frac{\partial}{\partial t}+X_h)(\mu(u))
\end{array}
\]
wich implies that $(\frac{\partial}{\partial t}+X_h)$ and  $(\frac{\partial}{\partial t}+X_h)^{\gamma}$ are $\mu\circ\gamma$-related.
\hfill $\Box$

\subsection{External Forces}

In this section we will apply the above general scheme to time-dependent systems and systems with external forces (see \cite{godbillon, cantrijn}).

A force is represented by a semi-basic $1$-form $F(t, v_q)=\alpha_i(t, q, \dot{q})\, dq^i$, wich is equivalent to give a fibred mappping
\[
\xymatrix{
\mathbb{R} \times TQ\ar[r]^{F}\ar[d]_{Id_{\mathbb{R}}\times\tau_Q}& T^*Q \ar[d]^{\pi_Q}\\
\mathbb{R}\times Q\ar[r]^{pr_Q}&  Q
}
\]
(see \cite{godbillon} for details). Assuming that our dynamical system is described by a regular lagrangian $L: TQ\longrightarrow \R$ and the force $F$, then using the Legendre transformation $FL: TQ\rightarrow \R$ we can transport $F$ to the hamiltonian side and define $\tilde{F}=F\circ (\mathbb{F}L)^{-1}$.

We have
\[
\xymatrix{
\mathbb{R}\times T^*Q\ar[r]^{\tilde{F}}\ar[d]_{Id_{\mathbb{R}}\times\pi_Q}& T^*Q \ar[d]^{\pi_Q}\\
\mathbb{R}\times Q\ar[r]^{pr_Q}&  Q
}
\]
where $pr_Q(t,q)=q$

Given a hamiltonian $h:\mathbb{R}\times T^*Q\rightarrow \mathbb{R}$, then the evolution of the system with external force $\tilde{F}$ is now given by
\[
\frac{\partial}{\partial t}+X_h+V_{\tilde{F}}
\]
where $V_{\tilde{F}}$ is the vector field determined by
\[
V_{\tilde{F}}(t,\alpha_q)=\Lambda_Q^{\sharp}(\pi_{Q}^*({\tilde{F}}(t,\alpha_Q))),
\]
$\Lambda_Q$ being the canonical Poisson structure on $T^*Q$.

In bundle coordinates $\frac{\partial}{\partial t}+X_h+V_{\tilde{F}}$ provides the differential equation
\[
\begin{array}{l}
\dot{q}_i=\displaystyle\frac{\partial h}{\partial p_ i}\\ \noalign{\medskip}
\dot{p}_i=-\displaystyle\frac{\partial h}{\partial q_i}-{\tilde{F}}_i.
\end{array}
\]

We can consider $T^*(\mathbb{R}\times Q)$ with the almost-Poisson structure $\tilde{\Lambda}$ given by $\tilde{\Lambda}=\Lambda_{\mathbb{R}\times Q}+V_F\wedge\frac{\partial}{\partial e} $ (recall the definition of $e$ in the previous section). In local coordinates
\[
\tilde{\Lambda}={\tilde{F}}_i\frac{\partial}{\partial e}\wedge\frac{\partial}{\partial p_i}+\frac{\partial}{\partial t}\wedge \frac{\partial}{\partial e}+ \frac{\partial}{\partial q^i}\wedge \frac{\partial}{\partial p_i}
\]
It is easy to see that the characteristic distribution of $\tilde{\Lambda}$ is the whole space (see \eqref{dd}).

We can define $h_{ext}=\mu^*h+e$, where $\mu$ is defined in the same way that in \ref{mu}. We can construct the hamiltonian vector field $X_{h_{ext}}=\tilde{\Lambda}^{\sharp}(dh_{ext})$. Due to the definition of $\tilde{\Lambda}$
it is easy to see that $\mu_*(X_{h_{ext}})=\frac{\partial}{\partial t}+X_h+V_{\tilde{F}}$.

The following diagram summarizes our construction
\[
\xymatrix{
& T^*(\mathbb{R}\times Q) \ar[dd]^{\pi_{\mathbb{R}\times Q}} \ar[dl]_{\mu} \\
\mathbb{R}\times T^*Q\ar[dr]^{\pi} \\
&\mathbb{R}\times Q
}
\]

If $\gamma$ is a section of $\pi_{\mathbb{R}\times Q}$ (a $1$-form on $\mathbb{R}\times Q$) we can consider the section of $\pi$ given by $\mu\circ\gamma$ and define the vector field $(\frac{\partial}{\partial t}+X_h+V_{\tilde{F}})^{\gamma}$ on $\mathbb{R}\times M$
\[
(\frac{\partial}{\partial t}+X_h+V_{\tilde{F}})^{\gamma}=T\pi\circ (\frac{\partial}{\partial t}+X_h+V_F)\circ (\mu\circ \gamma)
\]
and we can state the following.

\begin{theorem}If Im($\gamma$) is a lagrangian manifold in $\left((T^*(\mathbb{R}\times Q),\tilde{\Lambda}\right)$, then the following assertions are equivalent.
\begin{enumerate}
\item $(\frac{\partial}{\partial t}+X_h+V_{\tilde{F}})$ and  $(\frac{\partial}{\partial t}+X_h+V_{\tilde{F}})^{\gamma}$ are $\mu\circ\gamma$-related
\item $dh_{ext}\in T\textrm{Im}(\gamma)^{\circ}+\langle dt\rangle$
\end{enumerate}
\end{theorem}
{\bf Proof:}

The proof is analogous to that in Theorem \ref{teo}

\hfill $\Box$

Next, we shall characterize when a section $\gamma$ is lagrangian.

\begin{proposition}A $1$-form on $\mathbb{R}\times Q$ is lagrangian with respect to $\tilde{\Lambda}$ if and only if 
\[
d\gamma=({\tilde{F}}\circ\mu\circ\gamma)\wedge dt
\]
\end{proposition}
{\bf Proof:}
Using the local expression of $\tilde{\Lambda}$ we have
\begin{equation}\label{dd}
\begin{array}{l}
\tilde{\Lambda}(dt)=-\frac{\partial}{\partial e} \\ \noalign{\medskip}
\tilde{\Lambda}(de)= -F_i\frac{\partial}{\partial p_ i}+\frac{\partial}{\partial t}\\ \noalign{\medskip}
\tilde{\Lambda}(dq^i)=-\frac{\partial}{\partial p^i} \\ \noalign{\medskip}
\tilde{\Lambda}(dp_i)=\frac{\partial}{\partial q_i}+F_i\frac{\partial}{\partial e}.
\end{array}
\end{equation}

It is easy to see that $\tilde{\Lambda}^{\sharp}$ is an isomorphism, and so we can define the corresponding almost-symplectic structure $\tilde{\Omega}$, that is $(\tilde{\Lambda}^{\sharp})^{-1}=\Omega^{\flat}$, and thus
\begin{equation}
\begin{array}{l}
\tilde{\Omega}^{\flat}(\frac{\partial}{\partial t})=-{\tilde{F}}_idq^i+de \\ \noalign{\medskip}
\tilde{\Omega}^{\flat}(\frac{\partial}{\partial e})=-dt \\ \noalign{\medskip}
\tilde{\Omega}^{\flat}(\frac{\partial}{\partial q^i})=dp_i+{\tilde{F}}_idt \\ \noalign{\medskip}
\tilde{\Omega}^{\flat}(\frac{\partial}{\partial p_i})=-dq^i.
\end{array}
\end{equation}

So we can conclude that 
\[
\tilde{\Omega}=dq^i\wedge dp_i+ de+ {\tilde{F}}_idq_i\wedge dt
\]

The image of the $1$-form $\gamma$ will be lagrangian for $\tilde{\Omega}$ if and only if 
\[
\begin{array}{ll}
0&=\gamma^*(\tilde{\Omega})=\gamma^*(dq^i\wedge dp_i+dt\wedge de+ {\tilde{F}}_idq_i\wedge dt)\\ \noalign{\medskip}&=\gamma^*(dq^i\wedge dp_i+dt\wedge de)+\gamma^*( {\tilde{F}}_idq_i\wedge dt)\\ \noalign{\medskip}&=-d\gamma+({\tilde{F}}_i\circ\mu\circ\gamma)dq^i\wedge dt
\end{array}
\]
and the result follows.
\hfill $\Box$
\begin{remark}
{\rm Our result generalizes the Hamilton-Jacobi theorem derived in \cite{balseiro} for the case of linear forces and  time-dependent systems \cite{aff}.
}
\end{remark}

\begin{remark}
{\rm
The above discussion can be extended to a more general setting using similar arguments than in  preceding sections.
}
\end{remark}
\section*{Acknowledgments}
This work has been partially supported by MICINN (Spain) MTM \linebreak 2010-21186-C02-01, MTM 2009-08166-E, the European project IRSES-project ``Geomech-246981'' and the ICMAT Severo Ochoa project SEV-2011-0087.
M. Vaquero wishes to thank MINECO for a FPI-PhD Position.


\begin{thebibliography} {99}

\bibitem{AM} R. Abraham, J.E. Marsden:
{\it Foundations of Mechanics}. 2nd ed., Benjamin-Cummings, Reading
(Ma), 1978.

\bibitem{Arnold} V.I. Arnold:
{\it Mathematical methods of classical mechanics}.
Second edition. Graduate Texts in Mathematics, 60. Springer-Verlag, New York, 1989.

\bibitem{balseiro} P. Balseiro, J.C. Marrero, D. Mart\'in de Diego, E. Padr\'on: A unified framework for mechanics: Hamilton-Jacobi equation and applications. {\it Nonlinearity} {\bf 23} (2010), no. 8, 1887--1918.

\bibitem{bates} L. Bates, J. Sniatycki: Nonholonomic reduction. {\it Rep. Math. Phys.}  {\bf 32} (1) (1993), 99�-115.

\bibitem{cantrijn} F. Cantrijn:
Vector fields generating invariants for classical dissipative systems. 
{\it J. Math. Phys.} {\bf 23} (9) (1982),  1589–-1595. 

\bibitem{nonlinear} F. Cantrijn, M. de Le\'on, D. Mart{\'\i}n de Diego:
On almost-Poisson structures in nonholonomic mechanics.
{\it Nonlinearity} {\bf 12} (1999), 721--737.

\bibitem{cari} J.F. Cari\~nena, X. Gracia, G. Marmo, E. Mart{\'\i}nez, M. Mu\~noz-Lecanda,
N. Rom\'an-Roy: Geometric Hamilton-Jacobi theory. {\sl Int. J.
Geom. Meth. Mod. Phys.} {\bf 3} (7) (2006), 1417--1458.

\bibitem{cari2} J.F. Cari\~nena, X. Gracia, G. Marmo, E. Mart{\'\i}nez, M. Mu\~noz-Lecanda,
N. Rom\'an-Roy: Geometric Hamilton-Jacobi theory for nonholonomic dynamical systems. {\sl Int. J.
Geom. Meth. Mod. Phys.}  {\bf 7} 3 (2010), 431--454.

\bibitem{godbillon} C. Godbillon: {\sl G\'eom\'etrie diff\'erentielle et m\'ecanique analytique.} Hermann, Paris 1969 183 pp.

\bibitem{sosa} M. Leok, T. Ohsawa, D. Sosa: Hamilton-Jacobi Theory for Degenerate Lagrangian Systems with Holonomic and Nonholonomic Constraints. {\sl arXiv:1109.6056}.

\bibitem{lmm1} M. de Le\'on, D. Iglesias-Ponte, D. Mart{\'\i}n de Diego:
Towards a Hamilton-Jacobi theory for nonholonomic mechanical
systems. {\sl Journal of Physics A: Math. Gen.} (2008), no. 1, 015205, 14 pp.

\bibitem{lmm2}M. de Le\'on, J.C. Marrero, D. Mart{\'\i}n de Diego:
A geometric Hamilton-Jacobi theory for classical field theories. In: {\sl Variations, geometry and physics},
129--140, Nova Sci. Publ., New York, 2009

\bibitem{lmm3} M. de Le\'on, J.C. Marrero, D. Mart{\'\i}n de Diego:
Linear almost Poisson structures and Hamilton-Jacobi equation. Applications to nonholonomic mechanics.
{\sl J. Geom. Mech.} {\bf 2} 2 (2010), 159--198.

\bibitem{vilarino} M. de Le\'on, D. Mart{\'\i}n de Diego, J.C. Marrero, M. Salgado, S. Vilari\~no:
Hamilton-Jacobi theory in $k$-symplectic field theories. {\sl Int. J.
Geom. Meth. Mod. Phys.}  {\bf 7} no. 8 (2010), 1491–1507.

\bibitem{hjsingular1} M. de Le\'on, J. C. Marrero,  D. Mart{\'\i}n de Diego, M. Vaquero:
A Hamilton-Jacobi theory for singular lagrangian systems. {\it Preprint}.

\bibitem{hjsingular2} M. de Le\'on, D. Mart{\'\i}n de Diego, M. Vaquero:
A Hamilton-Jacobi theory for singular lagrangian systems in the Skinner and Rusk setting. To appear in {\sl Int. J.
Geom. Meth. Mod. Phys.} (2012).

\bibitem{lmv} M. de Le\'on, D. Mart{\'\i}n de Diego, M. Vaquero: A geometric
Hamilton-Jacobi theory for multisymplectic field theories. {\it In
preparation}.

\bibitem{deleonrodrigues} M. de Le\'on, P. R. Rodrigues:
{\it Methods of differential geometry in analytical mechanics}. North-Holland Mathematics Studies, 158. North-Holland Publishing Co., Amsterdam, 1989.

\bibitem{liber} P. Libermann, Ch.M- Marle: {\sl Symplectic Geometry and Analytical Mechanics}.
D. Reidel Publishing Co., Dordrecht, 1987.

\bibitem{aff} J.C. Marrero, D. Sosa: The Hamilton-Jacobi equation on Lie affgebroids. {\it Int. J. Geom. Methods Mod. Phys.} {\bf 3} (2006), no. 3, 605--622. 

\bibitem{bloch} T. Oshawa, A.M. Bloch: Nonholonomic Hamilton-Jacobi equations and integrability.
{\sl J. Geom. Mech.} {\bf 1} 4 (2009), 461--481.

\bibitem{rund} H. Rund: {\sl The Hamilton-Jacobi Theory in the Calculus of
Variations}. Hazell, Watson and Viney Ltd., Aylesbury,
Buckinghamshire, U.K. 1966.

\bibitem{Vais} I. Vaisman: {\sl Lectures on the geometry of Poisson manifolds}.
Progress in Mathematics, 118. Birkhäuser Verlag, Basel, 1994.

\bibitem{van} A. J. van der Schaft, B. M.  Maschke:
On the Hamiltonian formulation of nonholonomic mechanical systems.
{\it Rep. Math. Phys.} {\bf 34} (2) (1994), 225�233.


\end{thebibliography}
\end{document}